# Experimental Observation of the Trapped Rainbow

V.N. Smolyaninova [1)], I.I. Smolyaninov, A.V. Kildishev [3)], V. M. Shalaev [3)]

[1)] Department of Physics Astronomy and Geosciences, Towson University, 8000 York Rd., Towson, MD 21252 USA

[3)] Birck Nanotechnology Centre, School of Electrical and Computer Engineering, Purdue University, IN 47907, USA

**Abstract:** We report on the first experimental demonstration of the broadband "trapped rainbow" in the visible frequency range using an adiabatically tapered waveguide. Being a distinct case of the slow light phenomenon, the trapped rainbow effect could be applied to optical computing and signal processing, and to providing enhanced light-matter interactions.

The concept of a "trapped rainbow" has attracted considerable recent attention. According to various theoretical models, a specially designed metamaterial [1] or plasmonic [2,3] waveguide has the ability to slow down and stop light of different wavelengths at different spatial locations along the waveguide, which is extremely attractive for such applications as spectroscopy on a chip. In addition, being a special case of the slow light phenomenon [4], the trapped rainbow effect may be used in applications such as optical signal processing and enhanced light-matter interactions [5]. On the other hand, unlike the typical slow light schemes, the proposed theoretical trapped rainbow arrangements are extremely broadband, and can trap a true rainbow ranging from violet to red in the visible spectrum. Unfortunately, due to the necessity of complicated nanofabrication and the difficulty of producing broadband metamaterials, the trapped rainbow schemes had previously remained in the theoretical domain only.

In this communication we demonstrate an experimental realization of the broadband trapped rainbow effect which spans the 457-633 nm range of the visible spectrum. Similar to our recent demonstration of broadband cloaking [6], the metamaterial properties necessary for device fabrication were emulated using an adiabatically tapered waveguide geometry. A 4.5-mm diameter double convex glass lens was coated on one side with a 30-nm gold film. The lens was placed with the gold-coated side down on top of a flat glass slide coated with a 70-nm gold film (Fig.1A). The air gap between these surfaces has been used as an adiabatically changing waveguide. Light from a multi-wavelength argon ion laser (operating at λ=457 nm, 465 nm, 476 nm, 488 nm and 514 nm) and 633-nm light from a He-Ne laser were coupled to the waveguide via side illumination. This multi-line illumination produced the appearance of white light illuminating the waveguide (Fig.1B). Light propagation through the waveguide was imaged from the top using an optical microscope (Fig. 1C). Since the waveguide width at the entrance point is large, the air gap waveguide starts as a multi-mode waveguide. Gradual tapering of the waveguide leads to mode number reduction: the colored rings around the central circular dark area each represent a location where the group velocity of the n-th waveguide mode becomes zero. These locations are defined by $r_n = \sqrt{(n+1/2)R\lambda}$, where R is the lens radius [6]. Finally, the light in the waveguide is completely stopped at a distance $r = \sqrt{R\lambda/2}$ from the point of contact between the gold surfaces. The group velocity of the only remaining waveguide mode at this point is zero. This is consistent with the fact that the area around the point of contact appears dark in Fig. 1C. Since the stop radius depends on the light wavelength, different light colors stop at different locations inside the waveguide, which is quite obvious from Fig.1C. Thus, the visible light rainbow has been stopped and "trapped." To our knowledge, this is the first experimental demonstration of the broadband trapped rainbow effect in the visible frequency range. The same principle can be applied to any spectral range of interest.

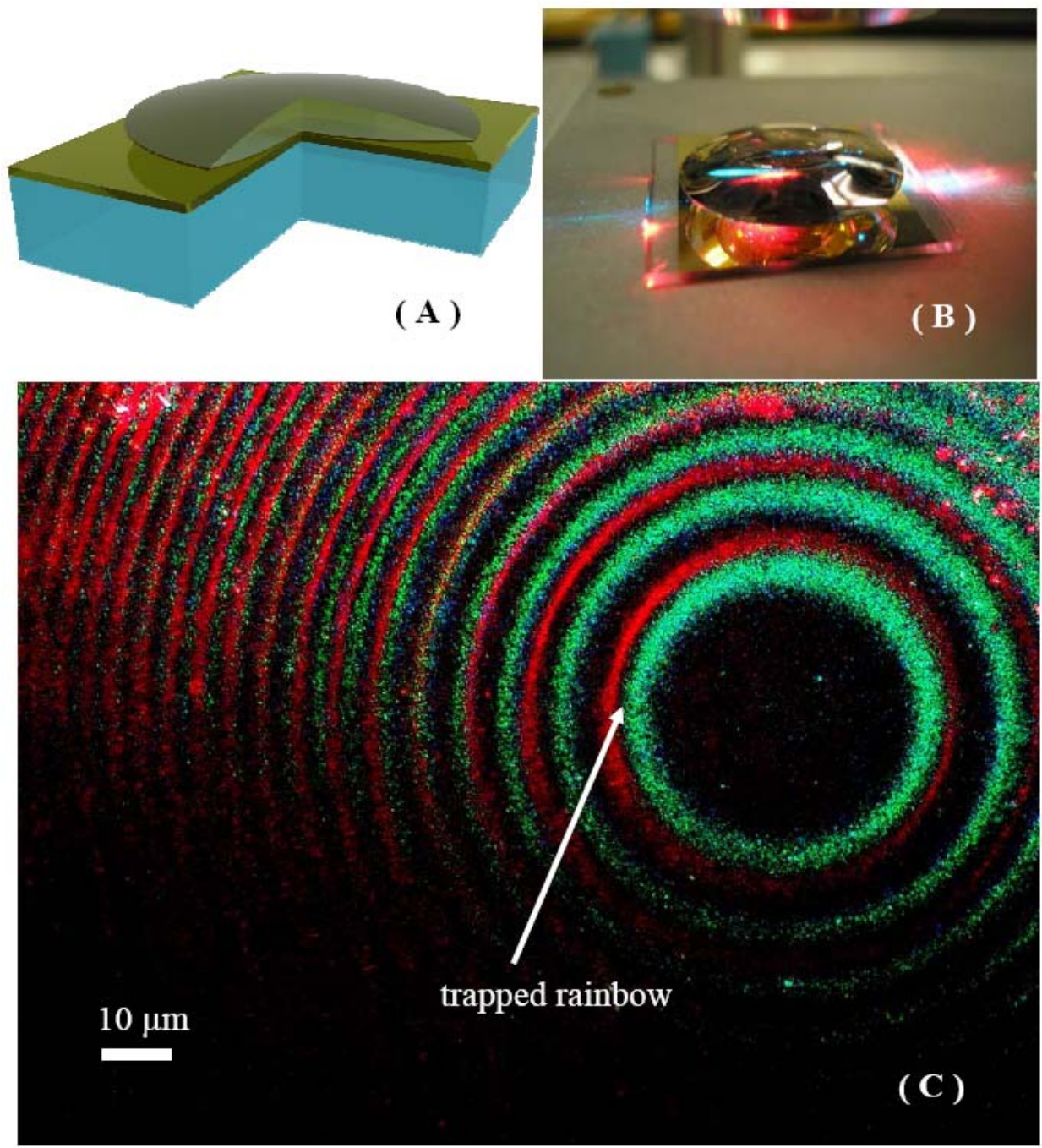

Figure 1: (A) Experimental geometry of the trapped rainbow experiment: a glass lens was coated on one side with a gold film. The lens was placed with the gold-coated side down on top of a flat glass slide also coated with a gold film. The air gap between these surfaces formed an adiabatically changing waveguide. (B) Photo of the trapped rainbow experiment: HeNe and Ar:Ion laser light is coupled into the waveguide. (C) Optical microscope image of the trapped rainbow.